\documentclass[mathleft]{an}
\usepackage{graphicx}
\usepackage{times}
\usepackage{amsmath}
\usepackage{amsfonts}
\usepackage{amssymb}

\begin{document}

\Pagespan{789}{}
\Yearpublication{2006}
\Yearsubmission{2005}
\Month{11}
\Volume{999}  
\Issue{88}
\DOI{This.is/not.aDOI} 

\title{BV photometry of a possible open star cluster pair NGC 7031/NGC 7086}

\author{V. S. Kopchev\thanks{Corresponding author: \email{kopchev@astro.bas.bg}\newline} \and  G. T. Petrov}

\titlerunning{BV photometry of a possible open star cluster pair NGC 7031/NGC 7086}
\authorrunning{V. S. Kopchev \& G. T. Petrov}
\institute{Institute of Astronomy, Bulgarian Academy of Sciences, 72 Tsarigradsko chaussee Blvd, 1784 Sofia, Bulgaria}

\received{?}
\accepted{?}
\publonline{?}


\abstract{
We present a CCD BV photometry of the possible binary open star cluster NGC 7031/NGC 7086.
The aim is to confirm or disprove their common nature on the grounds of their age and distance.
An age of 224 $\pm$ 25 Myr  and distance 831 $\pm$ 72 pc was determined for NGC 7031 and 
178 $\pm$ 25 Myr, 955 $\pm$ 84 pc for NGC 7086, respectively. Based on these differences in age and distance 
we conclude that the two clusters are most likely not formed together from one and the same 
Giant Molecular Cloud and thus are not a true binary cluster.
}

\maketitle

\section{Introduction}
A binary open star cluster is an object consisting  of two open clusters. They can be basically described as:
(i) binary physical systems with common origin formed together from one and the same Giant Molecular Cloud (GMC), 
having comparable ages and chemical compositions (we call this a true binary cluster);
(ii) binary physical systems arising from clusters formed in different parts of the Galaxy and forming a pair through gravitational capture. 
Cluster pairs formed through tidal  capture have different ages and/or \\chemical composition.

The existence of star cluster pairs in our neighbouring galaxies, the Magellanic Clouds, was confirmed by several authors (e.g. Bhatia \& Hatzidimitriou 1988; Bica et al.1992; Vallenari et al. 1998; Dieball \& Grebel 2000; de Oliveira et al. 2000). 
Dieball 2002 proposed a catalogue of binary and
multiple cluster candidates in the Large Magellanic Cloud with 473 members. There are more than 1600 open clusters in
our Galaxy but only one well established double or binary cluster namely $h + \chi$ Persei  (see e.g. Uribe et al. 2002, Kharchenko et al. 2005). 
Our Galaxy seems to show a lack of binary or multiple clusters as compared to the Magellanic Clouds.
It is still an open question why such objects do not exist in our Galaxy. 
The reasons could be that they have already merged or dissipated, or that they simply do not form.
Several lists of binary open cluster candidates have been proposed and studied by various authors:
Lyng\aa{} \& Wramdemark 1984; Pavlovskaya \& Filippova 1989; Tignanelli et al. 1990; Subramaniam et al. 1995;
Loktin 1997; Muminov 2000. One of the most complete and well studied lists is the one of Subramaniam et al. 1995 with
18 candidate pairs, including clusters NGC 7031 and NGC 7086 studied here.
Basic parameters of NGC 7031 and NGC 7086 as given in Dias et al. 2002 and WEBDA\footnote{http://www.univie.ac.at/webda} database are presented in Table 1.

\begin{table}
\caption{Basic parameters of the clusters NGC 7031 and NGC 7086.}
\begin{tabular}[h]{ccc}
\hline\noalign {\smallskip}
Parameter            	& NGC 7031   & NGC 7086    \\ 
\hline\noalign {\smallskip}
R.A.(2000)           	& 21:07:12	& 21:30:27    \\ 
Decl.(2000)          	& +50:52:30	& +51:36:00  \\
Distance (pc)        	& 900		& 1298           \\
Ang. diam (arcmin)   & 14.0 		& 12.0            \\
E(B-V) (mag)		& 0.854		& 0.807          \\ 
log(age)			& 8.138		 & 8.142         \\
\hline
\end{tabular}
\end{table}

Table 2 presents previous photographic studies of the open clusters NGC 7031 and NGC 7086.
The scatter, especially in the ages, is quite large for each cluster; all studies agree on a larger
distance to NGC 7086 in comparison with the distance proposed for NGC 7031.

For a good selection criterion for a binary cluster Dieball 2002 considers:
(i) the maximum centre-to-centre separation is $\approx$ 20 pc and
(ii) the age difference between the components of a binary cluster is either $\leq$ 10 Myr or their
ages agree well within the uncertainties of their age determination.
Based on N-body simulations Portegies Zwart \& Rusli 2007 conclude that a cluster pair with a smaller initial separation tends to merge in $\lesssim$ 60 Myr
due to loss of angular momentum from escaping stars, while clusters with a larger initial separation tend to become even more widely separated due to mass loss from the evolving stellar populations. This suggests that 20 pc is a good selection criterion.

The aim of our investigation is to determine more precisely cluster parameters such as reddening, distance and age using CCD photometry and applying criterion for binarity in order to confirm or disprove their binarity.

\begin{table*}
\caption{ Previous studies of the open clusters NGC 7031 and NGC 7086.}
\begin{tabular}[h]{ccccc}
\hline\noalign {\smallskip}
Name 	    & Citation 				     	& Distance & Age & E(B-V) \\
	  	    &       	   			     		&    pc        & Myr &  mag    \\
\hline\noalign {\smallskip}
NGC 7031   & Svolopoulos (1961)	     		& 760	  &        & 1.03  \\ 
	........ & Hoag et al. (1961)	    		& 900	  & 137 & 0.85	 \\
	........ & Lindoff (1968)		     		& 910 	  & 56   & 		 \\
	........ & Hassan \& Barbone (1973) 	& 710	  & 56   & 0.71	 \\
	........ & Ruprecht et al. (1981) 		&		  & 280 &		 \\
	........ & Janes \& Adler (1982)    		& 700	  &        & 0.71	 \\
NGC 7086   & Hassan (1967)		     		& 1170	  & 600 & 0.69	 \\
	........ & Lindoff (1968)		     		& 1205	  & 85   &		 \\
	........ & Janes \& Adler (1982)	     	& 1200	  &        & 0.70	 \\
\hline
\end{tabular}
\end{table*}

\section{Observations and data reduction}
The clusters were observed in the night of May 30 2006 with the 2-m Ritchey-Chretien telescope
of Rozhen National Astronomical Observatory - Bulgaria. The telescope is \\equipped with standard
Johnson filters and 1340x1300 VersArray 1300B CCD camera with 20 $\mu$m pixel size that
corresponds to 0.26 arcsec, giving a field of 5x6 arcmin\(^2\) in the sky.
The angular separation between the centres of the clusters is approximately 1$\degr$ and we cannot
observe them on a single frame. The cluster fields are shown in Fig. 1 and Fig. 2.
Table 3 presents an observation log, the seeing was between 1.6 - 1.9 arcsec.
Standard IRAF routines were used to reduce the data, photometry was carried out with DAOPHOT II.
The accuracy of the photometry is given in Fig. 3 where  colour and magnitude errors are plotted as a function of the V magnitude.
Instrumental magnitudes have been transformed to standard Johnson-Cousin system using a standard field around the cluster M92 (Majewski et al. 1994). The calibration equations we received are: 

b = B - (0.049 $\pm$ 0.006) + (0.28 $\pm$ 0.02)X\\

 - (0.114 $\pm$ 0.006)(B-V)\\

v = V - (0.737 $\pm$ 0.006) + (0.16 $\pm$ 0.02)X\\

 - (0.038 $\pm$ 0.006)(B-V),\\

where X is the airmass, the capital letters represent the standard magnitudes and colour
and lower-case letters denote instrumental magnitudes.

\begin{table}
\caption{Observation log.}
\begin{tabular}[h]{ccccc}
\hline\noalign {\smallskip}
Name	&Filter	& Exposure 	& Airmas \\
	  	&		&  (sec)      	&      	\\
\hline\noalign {\smallskip}
NGC 7031&	B	& 5 x 20  		& 1.10	\\ 
		&	V	& 5 x 10  		& 1.10	\\
\hline\noalign {\smallskip}
NGC 7086&	B	& 5 x 25		& 1.16	\\ 
		&	V	& 5 x 15   	& 1.16	\\
\hline
\end{tabular}
\end{table}

\begin{figure} 
\includegraphics[width=\columnwidth,angle=0]{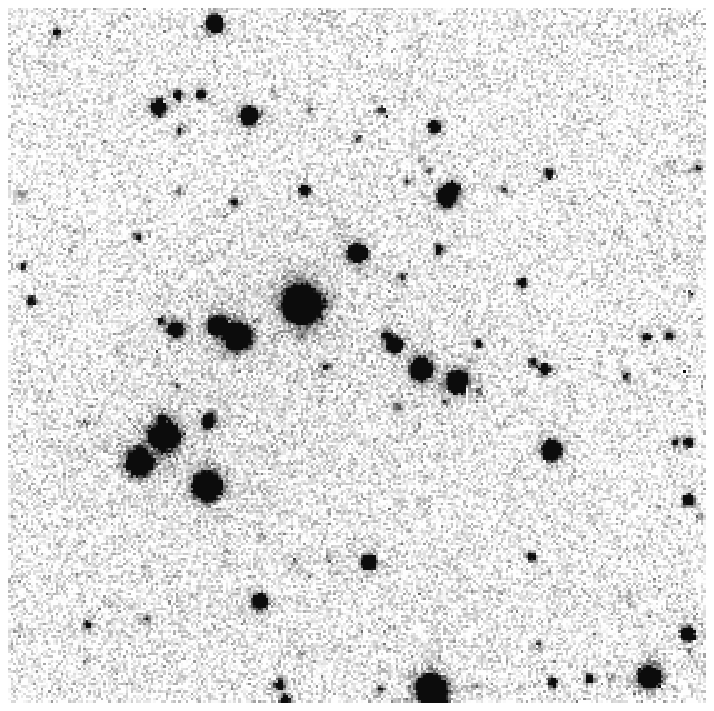}
\caption{$5^{\prime}$ x $6^{\prime}$ V band image of NGC 7031. North is to the top, east to the left.}
\end{figure} 

\begin{figure}
\includegraphics[width=\columnwidth,angle=0]{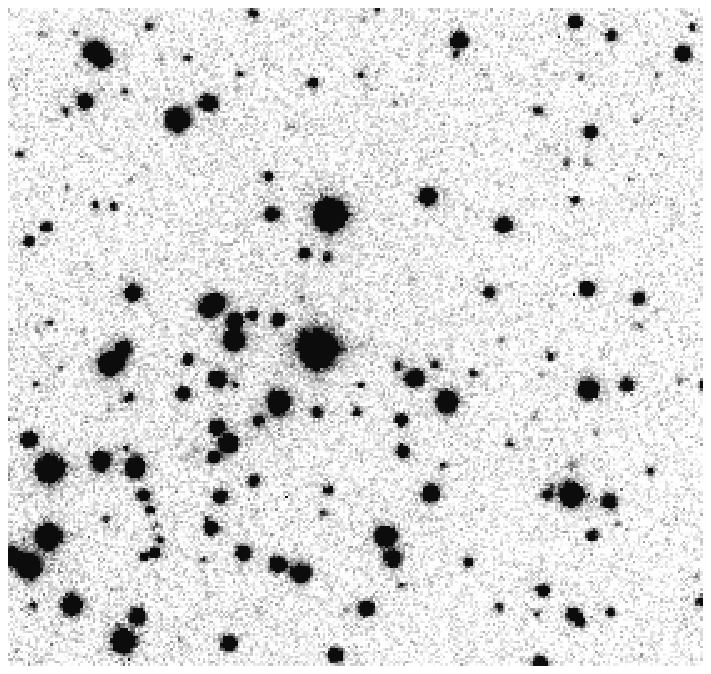}
\caption{$5^{\prime}$ x $6^{\prime}$ V band image of NGC 7086. North is to the top, east to the left.}
\end{figure} 

\begin{figure} 
\begin{tabular}{c}
\includegraphics[width=27mm,angle=-90]{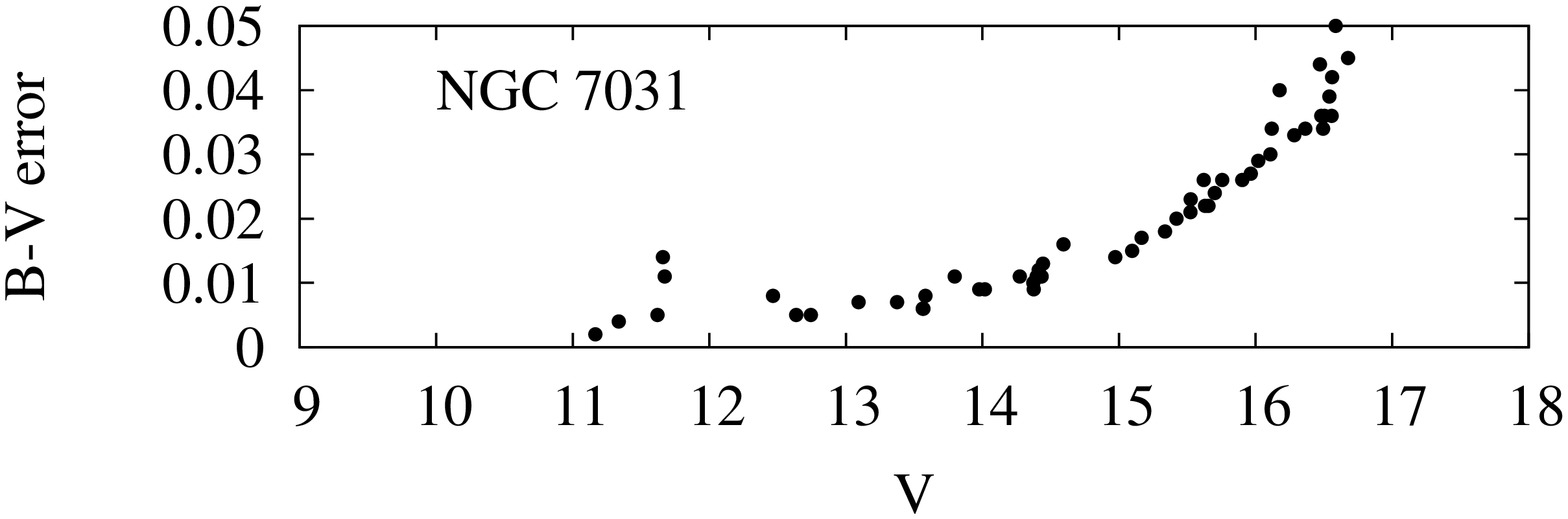}\\
\includegraphics[width=27mm,angle=-90]{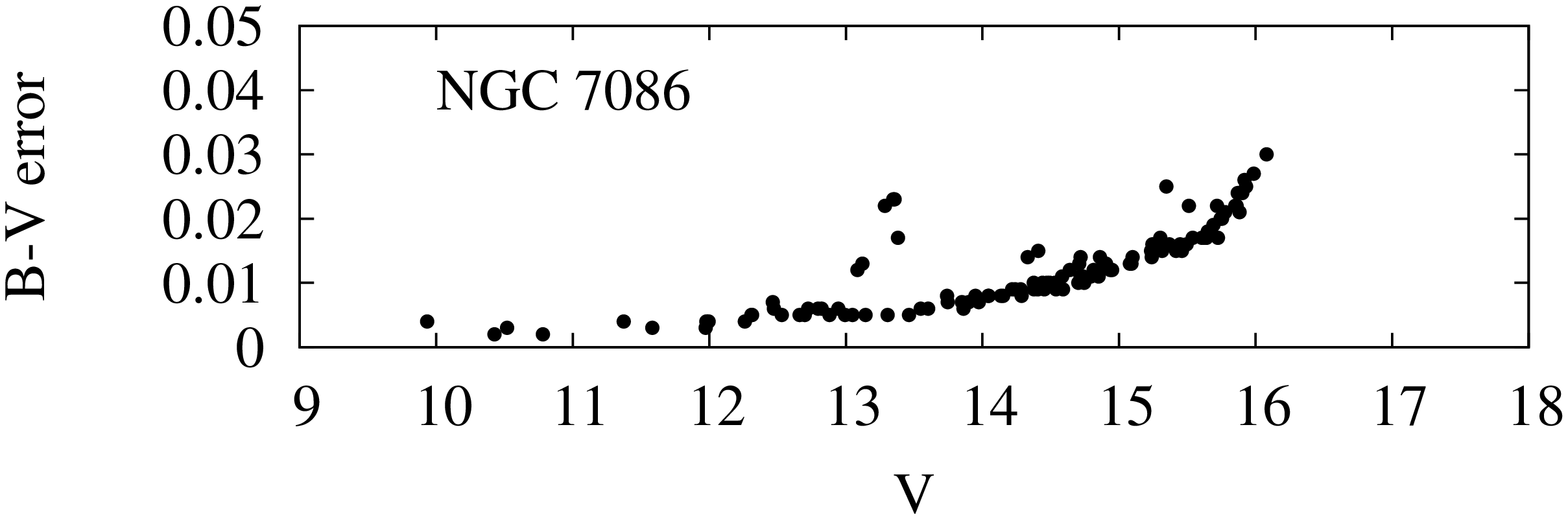}\\
\includegraphics[width=27mm,angle=-90]{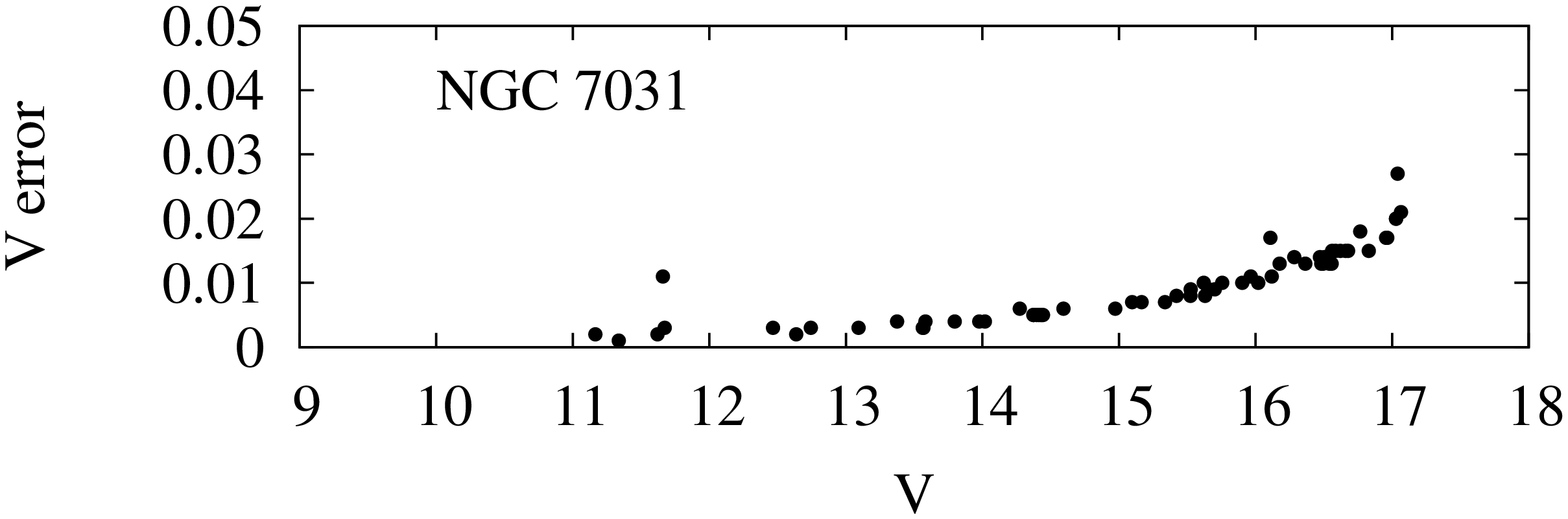}\\
\includegraphics[width=27mm,angle=-90]{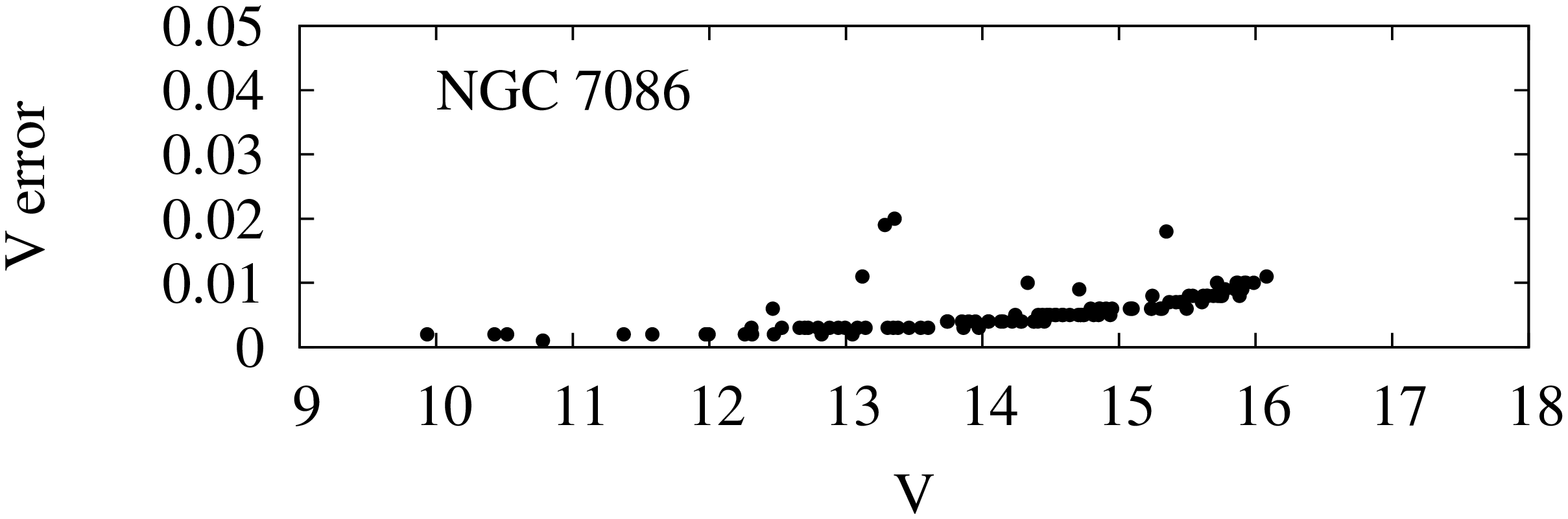}\\
 \end{tabular}
\caption{The color and magnitude errors from DAOPHOT as a function of the V magnitude.}
\end{figure} 

\section{Colour-Magnitude diagrams and cluster parameters}
The luminosity function of the clusters is presented in Fig. 4 with a dashed line for NGC 7031 and a solid line for NGC 7086.
Colour-magnitude diagrams (CMDs) are presented in Fig. 5 for NGC 7031 and Fig. 6 for NGC 7086,
where the solid line represents the zero age main sequence (ZAMS) taken from Schmidt-Kaler 1982.
Both CMDs reveal the presence of a reasonably broad and slightly evolved main sequence (MS),
typical of an early intermediate-age open cluster. The MS extends over a range of 6 magnitudes,
completeness limits are around V = 15.
To estimate the field star contamination (see, e.g. Bonatto \& Bica 2008) we used the Besancon model of stellar
population\footnote{http://bison.obs-besancon.fr/modele/}. The simulations determine
how many field stars can be expected in the fields of view of the clusters. For cluster NGC 7031 we do not expect
field star contamination in the range V = 11-12,  but for NGC 7086 the simulation showed one field star in the range V = 10 -11.
Distance module, reddenings and ages of the clusters have been derived by matching by eye the observed CMDs
to isochrones with Z = 0.020 from the Geneva group (Schaller et al. 1992), paying particular attention to the most likely 
shape of the main sequence, the turn-off point and the location of the evolved stars. 
In Table 4 we present our determination of cluster parameters.

\begin{figure}
\includegraphics[width=\columnwidth,angle=-90]{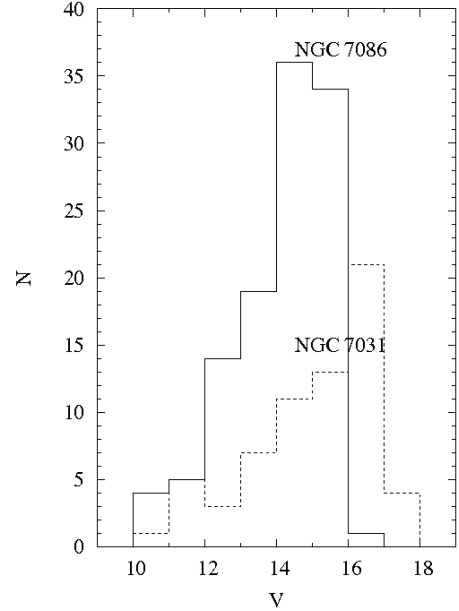}
\caption{Luminosity function of NGC 7031 and NGC 7086.}
\end{figure} 

\begin{figure} 
\includegraphics[width=\columnwidth,angle=-90]{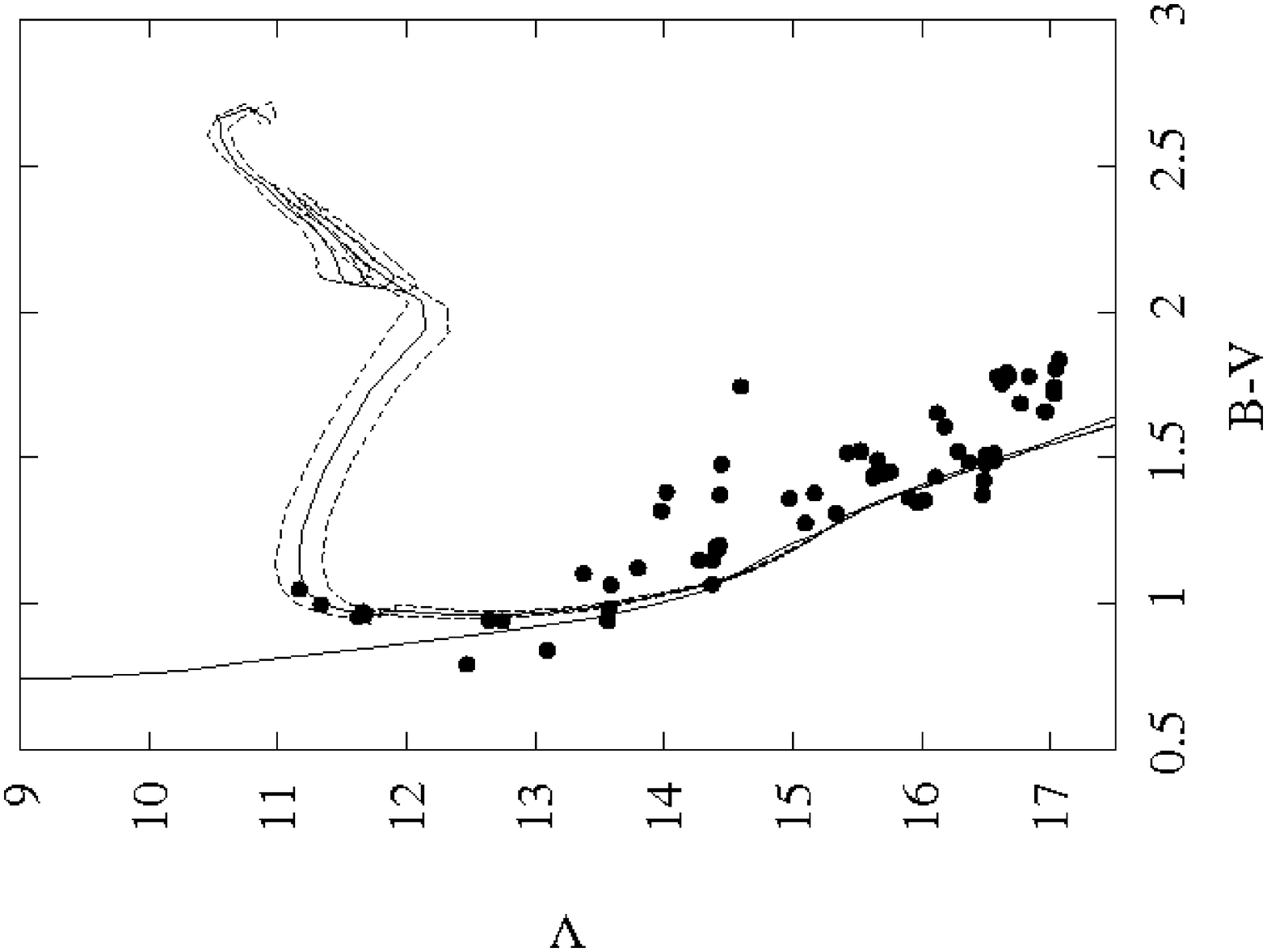}
\caption{The observed V, (B-V) diagram of NGC 7031 and the best isochrones fit with log(age) = 8.35 (solid curve).}
\end{figure} 

\begin{figure}
\includegraphics[width=\columnwidth,angle=-90]{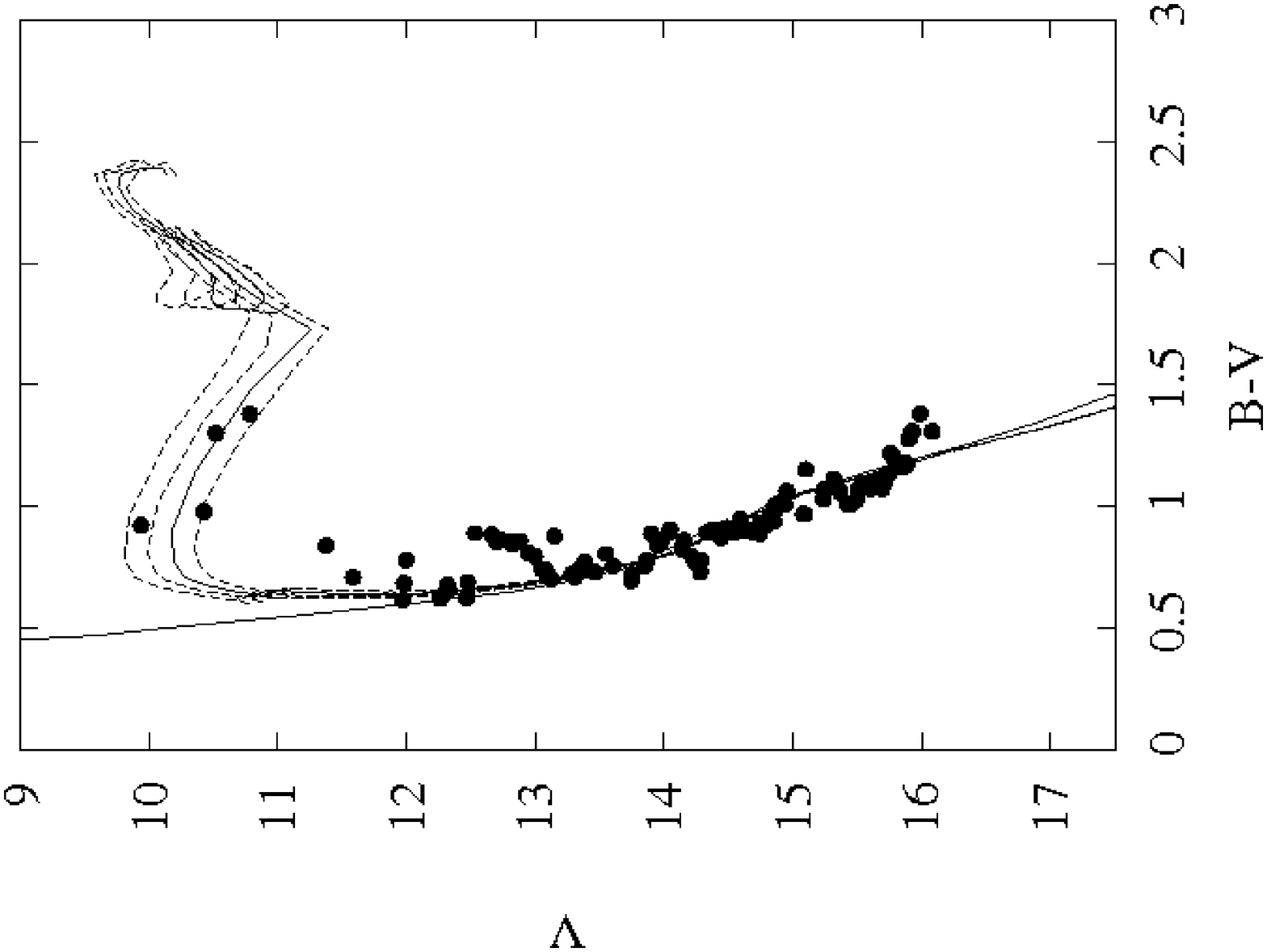}
\caption{The observed V, (B-V) diagram of NGC 7086 and the best isochrones fit with log(age) = 8.25 (solid curve).}
\end{figure} 

\begin{table}
\caption{Derived cluster parameters.}
\begin{tabular}[h]{ccc}
\hline\noalign {\smallskip}
Parameter            	& NGC 7031   	     & NGC 7086          \\ 
\hline\noalign {\smallskip}
E(B-V) (mag)         	& 1.05 $\pm$ 0.05 & 0.75 $\pm$ 0.05 \\ 
(m-M)v	          	& 9.6   $\pm$ 0.2  & 9.9   $\pm$ 0.2   \\
Distance (pc)        	& 831  $\pm$ 72   & 955 $\pm$ 84    \\
log(Age)			& 8.35 		     & 8.25 		   \\
Age (Myr)		& 224  $\pm$ 25   & 178 $\pm$ 25    \\ 
\hline
\end{tabular}
\end{table}

\section{Summary and conclusion}
We have presented BV CCD photometry of the closely projected open star clusters
NGC 7031 and NGC 7086. The results have been summarized in Table 4.
Our estimations of the age difference between NGC 7031 and NGC 7086, and difference in distance along the line of sight are 46 Myr and 124 pc, respectively. These results do not match with the criterion for a binary cluster and based on these results we
conclude that the two clusters are most likely not formed together from one and the same GMC and they are not a true binary cluster.

However, the ages of the two clusters roughly agree \\within the errors, and if we assume a mean distanse of  893 pc and an angular separation of 1$\degr$, this corresponds to a separation of 15 pc between the two clusters.These results are close enough for true binary. We caution that the distance differences  seem to indicate that the cluster separation is probably larger than 120 pc, thus it seems unlikely the two clusters form a true binary cluster, but it cannot be completely ruled out. Follow-up observations to determine the radial velocities of the clusters members and the motion of the clusters itself could put more light to this problem.

\acknowledgements
This research has made use of the WEBDA database, operated at the Institute
for Astronomy of the University of Vienna.


\begin{thebibliography}{}
\bibitem{} Bhatia, R. K., Hatzidimitriou, D.: 1988, MNRAS 230, 215
\bibitem{} Bica, E., Clari\'a, J. J., Dottori, H.: 1992, AJ 103, 1859
\bibitem{} Bonatto, C., Bica, E.: 2008, A \&Ap, 479, 741
\bibitem{} Dias, W. S., Alessi, B. S., Moitinho, A., L\`epine, J.R.D.: 2002, A\&A 389, 871
\bibitem{} Dieball, A.: 2002, "Binary star clusters in the Large Magellanic Cloud", Ph.D. Thesis, University of Bonn
\bibitem{} Dieball, A., Grebel, E. K.: 2000, A\&A 358, 897
\bibitem{} de Oliveira, M. R., Dutra, C. M., Bica, E., Dottori, H.: 2000, A\&AS 146, 57
\bibitem{} Hassan, S. M.: 1967, Z. Astrophysik 66, 6
\bibitem{} Hassan, S. H., Barbon, R.: 1973, MSAI 44, 39
\bibitem{} Hoag, A. A, Johnson, H. L, Iriarte, B., Mitchell, R. I, Hallam, K. L, Sharpless, S.: 1961, Publ. US. Nav. Obs. 17, 347
\bibitem{} Janes, K., Adler, D.: 1982, ApJS 49, 425
\bibitem{} Kharchenko, N. V., Piskunov, A. E., Roeser, S., Schilbach, E., Scholz, R.-D.: 2005, A\&Ap, 438, 1163
\bibitem{} Lindoff, U.: 1968, Arkiv f\"or Astronomi 5, 1
\bibitem{} Loktin, A. V.: 1997, Astronomical and Astrophysical Transactions 14, 181
\bibitem{} Lyng\aa{}, G., Wramdemark, S.: 1984, A\&A 132, 58
\bibitem{} Majewski, S. R., Kron, R. G., Koo, D. C., Bershady, M. A.: 1994, PASP 106, 1258
\bibitem{} Muminov, M.: 2000, Astronomische Gesellschaft Meeting Abstracts, Abstracts of Talks and Posters presented at the International Conference of the Astronomische Gesellschaft at Heidelberg, March 20-24, 2000, poster 70
\bibitem{} Pavlovskaya, E. D., Filippova, A. A.: 1989, Sov. Astron. 33, 602
\bibitem{} Portegies Zwart, S.F., Rusli, S.P.: 2007, MNRAS 374, 931
\bibitem{} Ruprecht, J., Balazs, B., White, R.: 1981, Budapest:Akademiai Kiado, "Catalogue of star clusters and associations"
\bibitem{} Schaller, G., Schaerer, D., Meynet, G., Maeder, A.: 1992, A\&AS 96, 269
\bibitem{} Schmidt-Kaler, T.S.,: 1982, Landolt-Bornstein Group 6, Vol.2b
\bibitem{} Subramaniam, A., Gorti, U., Sagar, R., Bhatt, H. C.: 1995, A\&A 302, 86
\bibitem{} Svolopoulos, S. N.: 1961, ApJ 134, 612
\bibitem{} Tignanelli, H., Vazquez, R. A, Mostaccio, C., Gordillo, S., Feinstein, A., Plastino A.: 1990, Rev. Mex. Astron. Astrofis. 21, 305
\bibitem{} Uribe, A., Garcia-Varela, J.-A., Sabogal-Martinez B.-E., Higuera, G. M. A., Brieva, E.: 2002, PASP, 114, 233
\bibitem{} Valenari, C., Bettoni, D., Chiosi, C.: 1998, A\&A 331, 506
\end{thebibliography}
\end{document}